\documentclass[twocolumn,showpacs,preprintnumbers,amsmath,amssymb,prl,nofootinbib]{revtex4-1}

\usepackage{graphicx}
\usepackage{dcolumn}
\usepackage{bm}
\usepackage{color}

\begin{document}

\title{Fundamental Limit on the Efficiency of Single-Photon Generation Based on Cavity Quantum Electrodynamics}

\author{Hayato Goto,$^1$ Shota Mizukami,$^2$ Yuuki Tokunaga,$^3$ and Takao Aoki$^2$}
\affiliation{$^1$Frontier Research Laboratory, 
Corporate Research \& Development Center, 
Toshiba Corporation, Kawasaki, Kanagawa 212-8582, Japan
\\
$^2$Department of Applied Physics, Waseda University, 3-4-1 Okubo, Shinjuku, Tokyo 169-8555, Japan
\\
$^3$NTT Secure Platform Laboratories, NTT Corporation, Musashino 180-8585, Japan}

\begin{abstract}
\noindent We analytically derive the upper bound on the overall efficiency 
of single-photon generation
based on cavity quantum electrodynamics (QED), 
where cavity internal loss is treated explicitly.
The internal loss leads to a tradeoff relation between 
the internal generation efficiency and the escape efficiency,
which results in a fundamental limit on the overall efficiency.
The corresponding lower bound on the failure probability is 
expressed only with an ``internal cooperativity," 
introduced here as the cooperativity parameter 
with respect to the cavity internal loss rate.
The lower bound is obtained by optimizing the cavity external loss rate,
which can be experimentally controlled 
by designing or tuning the transmissivity of the output coupler.
The model used here is general enough to treat
various cavity-QED effects,
such as the Purcell effect, 
on-resonant or off-resonant cavity-enhanced Raman scattering,
and vacuum-stimulated Raman adiabatic passage.
A repumping process, where the atom is reused after its decay to the initial ground state, 
is also discussed.
\end{abstract}

\maketitle

\textit{Introduction}.
Single-photon sources are a key component for photonic quantum information processing and quantum networking
\cite{Kimble2008a}. 
Single-photon sources based on cavity quantum electrodynamics (QED)
\cite{Eisaman2011a,Rempe2015a,Kuhn2010a,Law1997a,Vasilev2010a,Maurer2004a,Barros2009a,Kuhn1999a,Duan2003a}
are particularly promising, 
because they enable deterministic emission into a single mode, 
which is desirable for low-loss and scalable implementations. 
Many single-photon generation schemes
have been proposed and studied
using various cavity-QED effects,
such as the Purcell effect
\cite{Eisaman2011a,Rempe2015a,Kuhn2010a},
on-resonant~\cite{Kuhn2010a,Law1997a,Vasilev2010a} or 
off-resonant~\cite{Maurer2004a,Barros2009a} 
cavity-enhanced Raman scattering, and 
vacuum-stimulated Raman adiabatic passage (vSTIRAP)
\cite{Eisaman2011a,Rempe2015a,Kuhn2010a,Vasilev2010a,Kuhn1999a,Duan2003a,Maurer2004a}.

The overall efficiency of single-photon generation
based on cavity QED is composed of two factors: 
the internal generation efficiency $\eta_{\mathrm{in}}$
(probability that a photon is generated inside the cavity) and 
the escape efficiency $\eta_{\mathrm{esc}}$
(probability that a generated photon is extracted to a desired external mode).
The upper bounds on $\eta_{\mathrm{in}}$, based on the cooperativity parameter $C$~\cite{Rempe2015a},
have been derived for some of the above schemes
\cite{Rempe2015a,Kuhn2010a,Law1997a,Vasilev2010a}.
$C$ is inversely proportional to the total cavity loss rate, 
$\kappa=\kappa_{\mathrm{ex}}+\kappa_{\mathrm{in}}$,
where $\kappa_{\mathrm{ex}}$ and 
$\kappa_{\mathrm{in}}$
are the external and internal loss rates, respectively~\cite{comment-loss}.
Note that $\kappa_{\mathrm{ex}}$ can be experimentally controlled
by designing or tuning the transmissivity of the output coupler.
Thus, $\eta_{\mathrm{in}}$
is maximized by setting $\kappa_{\mathrm{ex}}$ to a small value
so that $\kappa \approx \kappa_{\mathrm{in}}$.
However, a low $\kappa_{\mathrm{ex}}$ results in a low escape efficiency $\eta_{\mathrm{esc}}=\kappa_{\mathrm{ex}}/\kappa$, which limits the channelling of the generated photons into the desired mode.
There is therefore a \textit{tradeoff} relation between $\eta_{\mathrm{in}}$ and $\eta_{\mathrm{esc}}$
with respect to $\kappa_{\mathrm{ex}}$, and $\kappa_{\mathrm{ex}}$ should be optimized
to maximize the overall efficiency.
This tradeoff relation has not been examined in previous studies,
where the internal loss rate
$\kappa_{\mathrm{in}}$
has not been treated explicitly. 
Additionally, previous studies on the photon-generation efficiency
have not taken account of
a repumping process, 
where the atom decays to the initial ground state via spontaneous emission
and is ``reused" for cavity-photon generation
\cite{Barros2009a}.

\begin{figure}[b]
	\includegraphics[width=0.5 \textwidth]{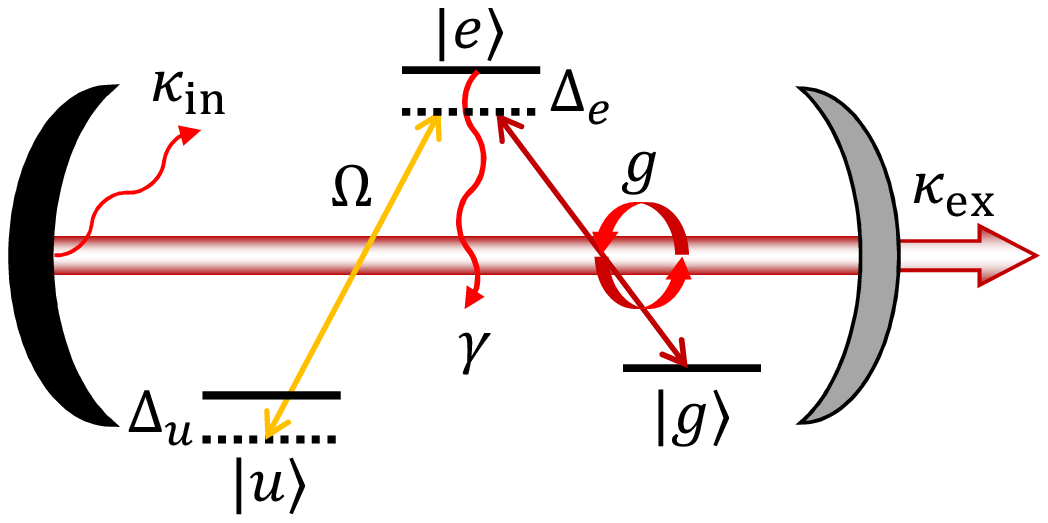}
	\caption{Cavity-QED system for single-photon generation.
	The atom is initially prepared in $|u\rangle$.
	$\kappa_{\mathrm{in}}$ and $\kappa_{\mathrm{ex}}$: 
	cavity internal and external loss rates, respectively.
	$g$: atom-cavity coupling rate via the $|g\rangle$-$|e\rangle$ transition.
	$\Omega$: Rabi frequency of the external field for the $|u\rangle$-$|e\rangle$ transition.
	$\Delta_e$ and $\Delta_u$: one-photon and two-photon detunings, respectively.
	$\gamma$: atomic decay rate due to spontaneous emission.}
	\label{fig-system}
\end{figure}

In this paper, we analytically derive the upper bound
on the overall efficiency of single-photon generation based
on cavity QED, 
by taking into account both the cavity
internal loss and the repumping process. 
We use the model shown in Fig.~\ref{fig-system}, 
which is able to describe most
of the previously proposed generation schemes, with or without the repumping process, in a unified and generalized manner.

In particular, we show that the lower
bound on the failure probability for single-photon generation, $P_F$, for the case of no repumping, is given by
\cite{comment-on-Goto,Goto2008a,Goto2010a}
\begin{align}
P_F
\ge
\frac{2}{\displaystyle 1+\sqrt{1+2C_{\mathrm{in}}}}
\approx \sqrt{\frac{2}{C_{\mathrm{in}}}},
\label{eq-PF}
\end{align}
where we have introduced the ``internal cooperativity,"
\begin{align}
C_{\mathrm{in}}=
\frac{g^2}{2\kappa_{\mathrm{in}} \gamma },
\label{eq-Cin}
\end{align}
as the cooperativity parameter with respect to $\kappa_{\mathrm{in}}$
instead of $\kappa$ for the standard definition, $C=g^2/(2\kappa \gamma )$
\cite{Rempe2015a}.
The approximation in Eq.~(\ref{eq-PF}) holds when $C_{\mathrm{in}} \gg 1$.

The lower bound on $P_F$ in Eq.~(\ref{eq-PF})
is obtained when $\kappa_{\mathrm{ex}}$ is set to its optimal value,
\begin{align}
\kappa_{\mathrm{ex}}^{\mathrm{opt}} 
\equiv \kappa_{\mathrm{in}} \sqrt{1+2C_{\mathrm{in}}},
\label{eq-optimal-kex}
\end{align}
and is simply expressed as 
$2\kappa_{\mathrm{in}}/\kappa^{\mathrm{opt}}$,
where
$\kappa^{\mathrm{opt}}
\equiv
\kappa_{\mathrm{in}}
+
\kappa_{\mathrm{ex}}^{\mathrm{opt}}$
\cite{comment-kex}.
Note that the experimental values of
$(g,\gamma,\kappa_{\mathrm{in}})$ 
determine which regime the system should be in:
the Purcell regime ($\kappa \gg g \gg \gamma$),
the strong-coupling regime ( $g \gg (\kappa, \gamma)$),
or the intermediate regime ($\kappa \approx g \gg \gamma$).

The remainder of this paper is organized as follows. 
First, we show that the present model is applicable to various cavity-QED single-photon generation schemes.
Next, we provide the basic equations for the present analysis.
Using these equations,
we analytically derive an upper bound on the success probability, $P_S=1-P_F$, of single-photon generation.
From here,
we optimize $\kappa_{\mathrm{ex}}$ and derive Ineq.~(\ref{eq-PF}).
We then briefly discuss the condition for typical optical cavity-QED systems.
Finally, the conclusion and outlook are presented.

\textit{Model}.
As shown in Fig.~\ref{fig-system}, we consider a cavity QED system with a $\Lambda$-type three-level atom in a one-sided cavity.
The atom is initially prepared in $|u\rangle$. 
The $|u\rangle$-$|e\rangle$ transition is driven with an external classical field, 
while the $|g\rangle$-$|e\rangle$ transition is coupled to the cavity.
This system is general enough to describe most of the cavity QED single-photon generation schemes.

For instance, by first exciting the atom to $|e\rangle$ with a resonant $\pi$ pulse 
(with time-dependent $\Omega$), or fast adiabatic passage (with time-dependent $\Delta_u$),
the atom is able to decay to $|g\rangle$ with a decay rate enhanced by the Purcell effect
\cite{Purcell1946a},
generating a single photon. 
Here, the Purcell regime is assumed. 
\cite{Rempe2015a,Kuhn2010a,Eisaman2011a}.

Another example is 
where the atom is weakly excited with small $\Omega$ and a cavity photon is generated by cavity-enhanced Raman scattering.
Here, $\kappa \gg g$ is assumed in the on-resonant case ($\Delta_e=\Delta_u=0$)
\cite{Kuhn2010a,Law1997a,Vasilev2010a}, while 
$\Delta_e \gg g$ is assumed in the off-resonant case ($\Delta_u=0$)
\cite{Barros2009a,Maurer2004a}.

A third example is based on 
vSTIRAP
\cite{Rempe2015a,Eisaman2011a,Kuhn2010a,Vasilev2010a,Kuhn1999a,Duan2003a,Maurer2004a},
where $\Omega$ is gradually increased, and where
the strong-coupling regime [$g \gg (\kappa, \gamma)$] and small detunings
($|\Delta_e|, |\Delta_u| \ll g$) are assumed.

\textit{Basic equations}.
The starting point of our study is the following master equation describing the cavity-QED system:
\begin{align}
\dot{\rho}
=&
\mathcal{L} \rho, ~
\mathcal{L}
=\mathcal{L}_{\mathcal{H}} + \mathcal{J}_u + \mathcal{J}_g + \mathcal{J}_o 
+ \mathcal{J}_{\mathrm{ex}} + \mathcal{J}_{\mathrm{in}},
\label{eq-master}
\\
\mathcal{L}_{\mathcal{H}} \rho
=&
-\frac{i}{\hbar} \left( \mathcal{H} \rho - \rho \mathcal{H}^{\dagger} \right),~
\mathcal{H}=H -i\hbar \left( \gamma \sigma_{e,e} + \kappa a^{\dagger} a \right),
\nonumber
\\
H
=&
\hbar \Delta_e \sigma_{e,e} + \hbar \Delta_u \sigma_{u,u}
\nonumber
\\
&+
i\hbar \Omega (\sigma_{e,u} - \sigma_{u,e} )
+
i\hbar g (a \sigma_{e,g} - a^{\dagger} \sigma_{g,e} ),
\label{eq-Hamiltonian}
\\
\mathcal{J}_{u} \rho
=&
2 \gamma r_u \sigma_{u,e} \rho \sigma_{e,u},~
\mathcal{J}_{g} \rho
=
2 \gamma r_g \sigma_{g,e} \rho \sigma_{e,g},
\nonumber
\\
\mathcal{J}_{o} \rho
=&
2 \gamma r_o \sigma_{o,e} \rho \sigma_{e,o},~
\mathcal{J}_{\mathrm{ex}} \rho
=
2 \kappa_{\mathrm{ex}} a \rho a^{\dagger},~
\mathcal{J}_{\mathrm{in}} \rho
=
2 \kappa_{\mathrm{in}} a \rho a^{\dagger},
\nonumber
\end{align}
where $\rho$ is the density operator describing the state of the system;
the dot denotes differentiation with respect to time;
$H$ is the Hamiltonian for the cavity-QED system;
$a$ and $a^{\dagger}$ are respectively the annihilation and creation operators for cavity photons; 
$|o\rangle$ is, if it exists, a ground state other than $|u\rangle$ and $|g\rangle$;
$r_u$, $r_g$, and ${r_o=1-r_u-r_g}$ are respectively the branching ratios 
for spontaneous emission from $|e\rangle$ to $|u\rangle$, $|g\rangle$, and $|o\rangle$; and
$\sigma_{j,l}=|j\rangle \langle l|$ ($j,l=u, g, e, o$) are atomic operators.
In the present work, we assume no pure dephasing
\cite{comment-dephasing}.

\begin{figure}[b]
	\includegraphics[width=0.5 \textwidth]{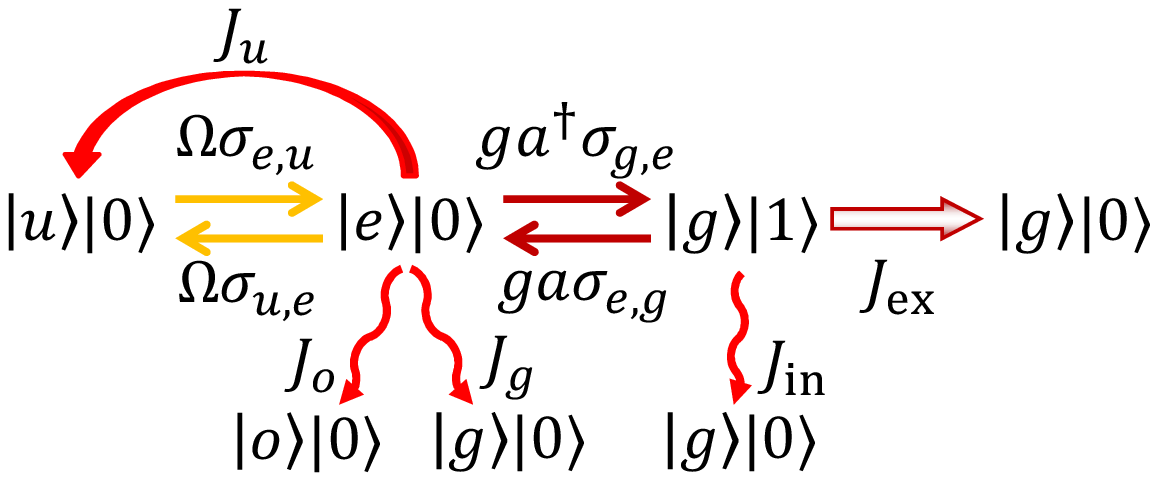}
	\caption{Transitions in Eqs.~(\ref{eq-master}) and (\ref{eq-Hamiltonian}).}
	\label{fig-transition}
\end{figure}

The transitions corresponding to the terms in Eqs.~(\ref{eq-master}) and (\ref{eq-Hamiltonian}) 
are depicted in Fig.~\ref{fig-transition}, where
the second ket vectors denote cavity photon number states.
Once the state of the system becomes $|g\rangle |0\rangle$ or $|o\rangle |0\rangle$
by quantum jumps,
the time evolution stops.
Among the quantum jumps,
$\mathcal{J}_{\mathrm{ex}}$
corresponds to the success case where a cavity photon is emitted into the external mode,
and the others result in failure of emission.
Taking this fact into account,
we obtain the following formal solution of the master equation
\cite{Carmichael}:
\begin{align}
\rho_c (t)
=&
\mathcal{V}_{\mathcal{H}}(t,0) \rho_0
+
\int_0^t \! dt' 
\mathcal{J}_{\mathrm{ex}}
\mathcal{V}_{\mathcal{H}} (t',0) \rho_0
\nonumber
\\
&+
\int_0^t \! dt' 
\mathcal{V}_c (t,t') 
\mathcal{J}_u
\mathcal{V}_{\mathcal{H}} (t',0) \rho_0,
\label{eq-rho-c}
\end{align}
where $\rho_c$ denotes the density operator conditioned on no quantum jumps of 
$\mathcal{J}_g$, $\mathcal{J}_o$, and $\mathcal{J}_{\mathrm{in}}$,
$\rho_0 = |u\rangle |0\rangle \langle u| \langle 0|$ is the initial density operator,
and $\mathcal{V}_{\mathcal{H}}$ and $\mathcal{V}_c$ are
the quantum dynamical semigroups defined as follows:
\begin{align}
\frac{d}{dt} \mathcal{V}_{\mathcal{H}}(t,t') = 
\mathcal{L}_{\mathcal{H}}(t) \mathcal{V}_{\mathcal{H}}(t,t'),~
\frac{d}{dt} \mathcal{V}_c(t,t') = \mathcal{L}_c(t) \mathcal{V}_c(t,t'),
\nonumber
\end{align}
where 
$\mathcal{L}_c
=\mathcal{L}_{\mathcal{H}} + \mathcal{J}_u + \mathcal{J}_{\mathrm{ex}}$
is the Liouville operator for the conditioned time evolution.
Note that $\rho_c(t) = \mathcal{V}_c(t,0) \rho_0$.
The trace of $\rho_c$ decreases from unity for ${t>0}$.
This decrease corresponds to the failure probability due to 
$\mathcal{J}_g$, $\mathcal{J}_o$, and $\mathcal{J}_{\mathrm{in}}$
\cite{Carmichael,Plenio1998a}.

Note that $\rho_{\mathcal{H}}(t)=\mathcal{V}_{\mathcal{H}}(t,0) \rho_0$ can be expressed with a state vector
as follows:
\begin{align}
\rho_{\mathcal{H}}(t)= |\psi (t) \rangle \langle \psi (t)|,~
i\hbar |\dot{\psi} \rangle = \mathcal{H} |\psi \rangle,~
|\psi (0) \rangle = |u \rangle |0 \rangle.
\nonumber
\end{align}
Setting $|\psi \rangle = \alpha_u |u \rangle |0 \rangle + \alpha_e |e \rangle |0 \rangle + \alpha_g |g \rangle |1 \rangle$,
the non-Hermitian Schr\"odinger equation is given by
\begin{align}
&
\dot{\alpha_u}=
-i\Delta_u \alpha_u
-\Omega \alpha_e,
\label{eq-alpha-u}
\\
&
\dot{\alpha_e}=
-(\gamma + i \Delta_e) \alpha_e + \Omega \alpha_u + g \alpha_g,
\label{eq-alpha-e}
\\
&
\dot{\alpha_g}=
-\kappa \alpha_g - g \alpha_e.
\label{eq-alpha-g}
\end{align}

Using the state vector and the amplitudes,
Eq.~(\ref{eq-rho-c}) becomes 
\begin{align}
\rho_c (t)
=&
|\psi (t) \rangle \langle \psi (t)|
+
2 \kappa_{\mathrm{ex}}
\int_0^t \! dt' 
|\alpha_g (t')|^2
|g \rangle |0 \rangle \langle g| \langle 0|
\nonumber
\\
&+
2 \gamma r_u
\int_0^t \! dt' 
|\alpha_e (t')|^2
\mathcal{V}_c (t,t') 
\rho_0.
\label{eq-rho-c-2}
\end{align}

\textit{Upper bound on success probability}. 
A successful photon generation and extraction event is defined by the condition that the final atom-cavity state is $|g\rangle|0\rangle$, 
and that the quantum jump $\mathcal{J}_{\mathrm{ex}}$ has occurred. 
The success probability, $P_S$, of the single-photon generation
is therefore formulated by $P_S = \langle g| \langle 0| \rho_c(T) |g \rangle |0 \rangle$ 
for a sufficiently long time $T$.
Using Eq.~(\ref{eq-rho-c-2}), we obtain
\begin{align}
P_S 
=&
2 \kappa_{\mathrm{ex}}
\int_0^T \! dt 
|\alpha_g (t)|^2
\nonumber
\\
&+
2 \gamma r_u
\int_0^T \! dt 
|\alpha_e (t)|^2
\langle g| \langle 0| 
\mathcal{V}_c (T,t) 
\rho_0
|g \rangle |0 \rangle.
\label{eq-PS-formula}
\end{align}

Here we assume the following inequality:
\begin{align}
\langle g| \langle 0| 
\mathcal{V}_c (T,t) 
\rho_0
|g \rangle |0 \rangle 
\le 
\langle g| \langle 0| 
\mathcal{V}_c (T,0) 
\rho_0
|g \rangle |0 \rangle = P_S.
\label{eq-Vc}
\end{align}
This assumption is natural because 
$\mathcal{V}_c (t,t')$ should be designed to maximize $P_S$
\cite{comment-Vc}.
Thus we obtain
\begin{align}
P_S 
\le
\frac{2 \kappa_{\mathrm{ex}} I_g}
{\displaystyle 1-2 \gamma r_u I_e},
\label{eq-PS-inequality}
\end{align}
where ${I_g = \int_0^T \! dt |\alpha_g (t)|^2}$
and ${I_e = \int_0^T \! dt |\alpha_e (t)|^2}$.

The two integrals, $I_g$ and $I_e$, can be evaluated as follows.
First, we have 
\begin{align}
\frac{d}{dt} \langle \psi |\psi \rangle
= -2\gamma |\alpha_e|^2 - 2\kappa |\alpha_g|^2
~\Rightarrow~ 
2\gamma I_e + 2\kappa I_g \approx 1,
\label{eq-norm}
\end{align}
where 
${\langle \psi (0)|\psi (0) \rangle =1}$ and
${\langle \psi (T)|\psi (T) \rangle \approx 0}$ have been used assuming a sufficiently long time $T$.
Next, using Eq.~(\ref{eq-alpha-g}),
we obtain
\begin{align}
&
I_e
= 
\int_0^T \! dt \frac{|\dot{\alpha_g}(t) + \kappa \alpha_g (t)|^2}{g^2}
\nonumber
\\
&=
\int_0^T \! dt \frac{|\dot{\alpha_g}(t)|^2 + \kappa^2 |\alpha_g (t)|^2}{g^2}
+
\frac{\kappa}{g^2}
\left[
|\alpha_g(T)|^2
-
|\alpha_g(0)|^2
\right]
\nonumber
\\
&\approx
\frac{I'_g}{g^2}
+\frac{\kappa^2}{g^2} I_g,
\label{eq-Ie}
\end{align}
where we have used ${|\alpha_g(0)|^2=0}$ and ${|\alpha_g(T)|^2\approx 0}$ and
have set ${I'_g = \int_0^T \! dt |\dot{\alpha}_g (t)|^2}$.
Using Eqs.~(\ref{eq-norm}) and (\ref{eq-Ie}),
we obtain
\begin{align}
I_g
&=
\frac{C}{\kappa (1+2C)}
\left(
1-
\frac{I'_g}{\kappa C}
\right),
\label{eq-Ig-result}
\\
I_e
&=
\frac{1}{2\gamma}
\left[
1-
\frac{2C}{1+2C}
\left(
1-
\frac{I'_g}{\kappa C}
\right)
\right].
\label{eq-Ie-result}
\end{align}

Substituting Eqs.~(\ref{eq-Ig-result}) and (\ref{eq-Ie-result}) 
into Ineq.~(\ref{eq-PS-inequality}),
the upper bound on $P_S$ is finally obtained as follows:
\begin{align}
P_S 
&\le
\frac{\kappa_{\mathrm{ex}}}{\kappa}
\frac{2C}{1+2C}
\frac{\displaystyle 1-\frac{I'_g}{\kappa C}}
{\displaystyle 1-r_u + r_u \frac{2C}{1+2C} \left( 1-\frac{I'_g}{\kappa C} \right)}
\nonumber
\\
&\le
\left( 1-
\frac{\kappa_{\mathrm{in}}}{\kappa} 
\right)
\left( 1-
\frac{1}{1+2C}
\right)
\sum_{n=0}^{\infty} 
\left(
\frac{r_u}{1+2C}
\right)^n,
\label{eq-PS}
\end{align}
where we have used $0\le 1 - I'_g/(\kappa C) \le 1$~\cite{comment-Ig}.
The equality approximately holds when
the system varies slowly and the following condition holds:
\begin{align}
\frac{1}{\kappa}
\int_0^T \! dt |\dot{\alpha_g}(t)|^2
\ll C.
\end{align}

The upper bound on the success probability given by Ineq.~(\ref{eq-PS}) 
is a unified and generalized version of previous results~\cite{Kuhn2010a,Law1997a,Vasilev2010a,comment-storage,Gorshkov2007a,Dilley2012a}, 
which did not treat explicitly internal loss, detunings, or repumping.
The upper bound has a simple physical meaning.
The first factor is the escape efficiency $\eta_{\mathrm{esc}}$.
The product of the second and third factors is 
the internal generation efficiency $\eta_{\mathrm{in}}$.
Each term of the third factor represents the probability that 
the decay from $|e \rangle$ to $|u \rangle$ occurs $n$ times.
Note that $\eta_{\mathrm{in}}$ is increased by the repumping process.

So far, the photons generated by repumping after decay to $|u\rangle$ are counted,
as in some experiments
\cite{Barros2009a}.
However, such photons may have time delays or different pulse shapes 
from photons generated without repumping,
and are therefore not useful for some applications, such as photonic qubits.
If the photons generated by repumping are not counted,
we should consider the state conditioned further on no quantum jump of $\mathcal{J}_u$.
In this case, the upper bound on the success probability is obtained by
modifying Ineq.~(\ref{eq-PS}) with $r_u=0$.

The contribution of the repumping to $P_S$,
denoted by $P_{\mathrm{rep}}$, 
is given by the second term
in the right-hand side of Eq.~(\ref{eq-PS-formula}).
Using Eqs.~(\ref{eq-Ie-result}) and (\ref{eq-PS}),
we can derive an upper bound on $P_{\mathrm{rep}}$
as follows:
\begin{align}
P_{\mathrm{rep}}
\le
2\gamma r_u I_e P_S
&\le
\frac{\kappa_{\mathrm{ex}}}{\kappa} 
\frac{2C}{1+2C}
\sum_{n=1}^{\infty} 
\left(
\frac{r_u}{1+2C}
\right)^n
\nonumber
\\
&=
\frac{\kappa_{\mathrm{ex}}}{\kappa} 
\frac{2C}{1+2C}
\frac{r_u}{1+2C-r_u}.
\label{eq-Prepump}
\end{align}
Thus, the contribution of the repumping
is negligible when $C \gg 1$ or when $r_u \ll 1$.

\textit{Fundamental limit on single-photon generation based on cavity QED}.
The reciprocal of the upper bound on $P_S$
is simplified as
\begin{align}
\left(
1 + \frac{\kappa_{\mathrm{in}}}{\kappa_{\mathrm{ex}}}
\right)
\left[
1 + \frac{1-r_u}{2C_{\mathrm{in}}} 
\left(
1 + \frac{\kappa_{\mathrm{ex}}}{\kappa_{\mathrm{in}}}
\right)
\right].
\end{align}
This can be easily minimized with respect to $\kappa_{\mathrm{ex}}$,
which results in the following lower bound on $P_F$:
\begin{align}
P_F
\ge
\frac{2}{\displaystyle 1+\sqrt{1+2C_{\mathrm{in}}/(1-r_u)}},
\label{eq-PF-ru}
\end{align}
where the lower bound is obtained when $\kappa_{\mathrm{ex}}$ is set to
\begin{align}
\kappa_{\mathrm{ex}}^{\mathrm{opt}} 
\equiv \kappa_{\mathrm{in}} \sqrt{1+2C_{\mathrm{in}}/(1-r_u)}.
\label{eq-optimal-kex-ru}
\end{align}
In the case of no repumping,
Eqs.~(\ref{eq-PF-ru}) and (\ref{eq-optimal-kex-ru}) are modified by $r_u=0$.
This leads to Ineq.~(\ref{eq-PF}) and Eq.~(\ref{eq-optimal-kex}).

The approximate lower bound in Ineq.~(\ref{eq-PF})
can be derived more directly from Ineq.~(\ref{eq-PS}) (${r_u=0}$)
using 
the arithmetic-geometric mean inequality as follows:
\begin{align}
P_F
\ge
\frac{\kappa_{\mathrm{in}}}{\kappa}
+
\frac{1}{2C+1}
-\frac{\kappa_{\mathrm{in}}}{\kappa}
\frac{1}{2C+1}
\approx
\frac{\kappa_{\mathrm{in}}}{\kappa}
+
\frac{\kappa \gamma}{g^2}
\ge
\sqrt{\frac{2}{C_{\mathrm{in}}}},
\nonumber
\end{align}
where ${\kappa_{\mathrm{in}} \ll \kappa}$
and ${C \gg 1}$ have been assumed.
Note that $\kappa$ is cancelled out by multiplying the two terms
\cite{comment-arithmetic-geometric}.

\textit{Typical optical cavity-QED systems}. 
In optical cavity-QED systems
where a single atom or ion is coupled to a single cavity mode
\cite{Law1997a,Vasilev2010a,Barros2009a,Maurer2004a,Kuhn1999a,Duan2003a},
the cavity-QED parameters are expressed as follows~\cite{Rempe2015a}:
\begin{align}
g &=
\sqrt{\frac{\mu_{g,e}^2 \omega_{g,e}}{2\epsilon_0 \hbar A_{\mathrm{eff}} L}},
\label{eq-g}
\\
\kappa_{\mathrm{in}} &=
\frac{c}{2L} \alpha_{\mathrm{loss}},
\label{eq-kappa-in}
\\
r_g \gamma &=
\frac{\mu_{g,e}^2 \omega_{g,e}^3}{6 \pi \epsilon_0 \hbar c^3},
\label{eq-gamma}
\end{align}
where $\epsilon_0$ is the permittivity of vacuum,
$c$ is the speed of light in vacuum,
$\mu_{g,e}$ and $\omega_{g,e}$ are the dipole moment
and frequency of the $|g\rangle$-$|e\rangle$ transition, respectively,
$L$ is the cavity length,
$A_{\mathrm{eff}}$ is the effective cross-section area of the cavity mode at the atomic position,
and
$\alpha_{\mathrm{loss}}$ is the one-round-trip cavity internal loss.
Substituting Eqs.~(\ref{eq-g})--(\ref{eq-gamma}) into the definition of 
$C_{\mathrm{in}}$,
we obtain
\begin{align}
\frac{2C_{\mathrm{in}}}{1-r_u}
&=
\frac{1}{\alpha_{\mathrm{loss}}} 
\frac{1}{r_A} 
\frac{r_g}{1-r_u}
\le
\frac{1}{\alpha_{\mathrm{loss}}} 
\frac{1}{r_A}, 
\label{eq-Cin-formula}
\end{align}
where $\lambda = 2\pi c/\omega_{g,e}$ is the wavelength
corresponding to $\omega_{g,e}$, 
$r_A=A_{\mathrm{eff}}/\sigma$
is the ratio of the cavity-mode area to the atomic absorption cross section 
${\sigma = 3\lambda^2/(2\pi)}$,
and
the inequality comes from $r_g/(1-r_u) \le 1$.
(The equality holds when ${r_o=0}$.)
Note that the cavity length $L$ and the dipole moment $\mu_{g,e}$
are cancelled out.
From Ineq.~(\ref{eq-PF-ru}), it turns out that
the single-photon generation efficiency is limited only by
the one-round-trip internal loss, ${\alpha_{\mathrm{loss}}}$, 
and the area ratio, $r_A$, 
even when counting photons generated by repumping.

\textit{Conclusion and outlook}.
By analytically solving the master equation for a general cavity-QED model,
we have derived an upper bound on the success probability of single-photon generation based on cavity QED
in a unified way.
We have taken cavity internal loss into account,
which results in a tradeoff relation between the internal generation efficiency and
the escape efficiency
with respect to the cavity external loss rate $\kappa_{\mathrm{ex}}$.
By optimizing $\kappa_{\mathrm{ex}}$,
we have derived a lower bound on the failure probability.
The lower bound is inversely proportional to 
the square root of the internal cooperativity $C_{\mathrm{in}}$.
This gives the fundamental limit of single-photon generation efficiency based on cavity QED.
The optimal value of $\kappa_{\mathrm{ex}}$ has also been given explicitly.
The repumping process, where  the atom decays to the initial 
ground state via spontaneous emission and is reused for 
cavity-photon generation has also been taken into account.

For typical optical cavity-QED systems,
the lower bound is determined only by
the one-round-trip internal loss and 
the ratio between the cavity-mode area and the atomic absorption cross section.
This result holds even when the photons generated by repumping are counted.

The lower bound is achieved in the limit that the variation of the system is sufficiently slow.
When the short generation time is desirable,
optimization of the control parameters will be necessary.
This problem is left for future work.

\section*{Acknowledgments}

The authors thank Kazuki Koshino, Donald White and Samuel Ruddell for their useful comments.
This work was supported by JST CREST Grant Number JPMJCR1771, Japan.


\begin{thebibliography}{19}
\bibitem{comment-loss}
The external loss is due to the extraction of cavity photons to the desired external mode via transmission of the mirror,
while the internal loss is due to undesirable scattering and absorption inside the cavity.
\bibitem{comment-Vc}
If
$\langle g| \langle 0| 
\mathcal{V}_c (T,t) 
\rho_0
|g \rangle |0 \rangle 
>
\langle g| \langle 0| 
\mathcal{V}_c (T,0) 
\rho_0
|g \rangle |0 \rangle$, then 
we should use $\mathcal{V}_c (T,t)$
for the single-photon generation,
instead of $\mathcal{V}_c (T,0)$.
\bibitem{comment-Ig}
Note that $I'_g \ge 0$ by definition and 
$1 - I'_g/(\kappa C) \ge 0$ because $I_g \ge 0$, by definition, in Eq.~(\ref{eq-Ig-result}).
\end{thebibliography}
\end{document}